\documentclass[a4paper]{article}

\usepackage{a4wide}
\usepackage{psfrag} 
\usepackage{enumerate}
\usepackage{mathtools}

\usepackage{graphicx}

\usepackage{multicol}

\usepackage{amsmath}
\usepackage{amsthm}
\usepackage{amsfonts}
\usepackage{amssymb}
\usepackage{mathrsfs}

\usepackage{lettrine}

\usepackage[sort]{cite}
\usepackage{times}
\usepackage{booktabs}

\usepackage{wasysym}

\usepackage{fancyhdr}

\newtheoremstyle{mystyle}{1pt}{1pt}{\normalfont}{\parindent}{\bfseries}{}{1em}{}
\theoremstyle{mystyle}
\newtheorem{Thm}{Theorem}

\newtheorem{Pro}{Proposition}
\newtheorem{Def}{Definition}
\newtheorem{Lem}{Lemma}
\newtheorem{Col}{Corollary}

\begin{document}                          

\title{Pole Placement Approach to Coherent Passive Reservoir
  Engineering for Storing Quantum
  Information\thanks{This research is supported by grants CE110001027 from
Australian Research Council Centre of Excellence for Quantum Computation
and Communication Technology, FA2386-12-1-4084 from AFOSR, and DP140101779
from Australian Research Council respectively. N.H.A. acknowledges CNRS funding 
under the JCJC INS2I 2016 ``QIGR3CF'' and JCNC INS2I 2017 ``QFCCQI'' projects.}}

\author{T.~Nguyen\thanks{ ARC Centre for Quantum Computation and
    Communication Technology, Research School of Engineering, The
    Australian National University, Canberra, ACT 2601, Australia.}
\and
Z.~Miao\thanks{ QUANTIC lab, INRIA Paris, 2 rue Simone Iff, 75012 Paris, France}
\and
Y.~Pan\thanks{Institute of Cyber-Systems and Control, Zhejiang University,
  Hangzhou 310027, China}
\and 
N.~H.~Amini\thanks{CNRS, Laboratoire des signaux et syst{\` e}mes (L2S),
  CentraleSup{\' e}lec, 3 rue Joliot Curie, 91192 Gif-Sur-Yvette, France}
\and
V.~Ugrinovskii\thanks{Corresponding author, School of Information Technology and Electrical
  Engineering, 
  University of New South Wales at the Australian Defence Force Academy,
  Canberra, ACT 2600, Australia}
\and
M.~R.~James$^\dagger$}

\maketitle

\begin{abstract}
Reservoir engineering is the term used in quantum control and
  information 
technologies to describe manipulating the environment within which an open
quantum system operates. Reservoir engineering is essential in
applications where storing quantum information is required. From the
control theory perspective, a quantum system is 
capable of storing quantum information if it possesses a so-called
decoherence free subsystem (DFS). This paper explores pole placement
techniques to facilitate synthesis of decoherence free subsystems via
coherent quantum feedback control. We discuss limitations of the
conventional `open loop' approach and propose a constructive feedback
design methodology for decoherence free subsystem engineering.  
It captures a quite general dynamic coherent feedback structure which allows
systems with decoherence free modes to be synthesized from components which
do not have such modes.

\end{abstract}

\section{Introduction}   \label{sec:intro}

The environment within which the quantum system operates typically has a
continuous degrading effect on the evolution of quantum particles. This effect
known as \emph{decoherence} is the reason for the continuous process of
degeneration of distinctly quantum states into classical
ones~\cite{Poyatos1996}.  
On the other hand, when a quantum system possesses
a subsystem isolated from the detrimental influence of the environment and
probing fields, the quantum information associated with dynamics of such a
system is preserved and can be used for quantum computation when
needed. In a sense, decoherence free subsystems (DFS) can play roles of
memory elements in quantum information processing. This  has motivated significant interest in the synthesis of quantum systems with a desired DFS
structure.

The problem of DFS synthesis has been found to be nontrivial --- it has
been shown in~\cite{Yamamoto2014} that conventional measurement feedback is
ineffective in producing quantum systems having a DFS, however certain
coherent controllers can overcome this limitation of the measurement-based
feedback controllers. The objective of this paper is to put this
observation on a solid systematic footing, by developing a quite general
constructive coherent synthesis procedure for generating quantum systems
with a DFS of desired dimension.

Our particular interest is in a class of quantum linear systems 
whose dynamics in the Heisenberg picture are described by complex quantum
stochastic differential equations expressed in terms of annihilation
operators only. Such systems are known to be passive~\cite{James2010}. Passivity
ensures that the system does not generate energy.  In addition, in such systems the
notion of system controllability by noise and that of observability from the
output field are known to be equivalent~\cite{Gough2015}. Also, one can readily
identify uncontrollable and unobservable subspaces of the passive system by
analyzing the system in the Heisenberg picture~\cite{Yamamoto2014}. These additional features of
annihilation only passive systems facilitate the task of synthesizing
decoherence free subsystems by means of coherent feedback. 

The focus on a general coherent feedback synthesis is the main distinct
feature of this paper which differentiates it from other works  of a
similar kind, notably from~\cite{Yamamoto2014,Nurdin2015}. The paper~\cite{Yamamoto2014}
presents an analysis of quantum systems equipped with coherent feedback for
the purpose of characterizing decoherence free subsystems, quantum nondemolished
(QND) variables and measurements capable of evading backaction;
in~\cite{Yamamoto2014} all these characteristics are expressed in geometric terms of
(un)controllable and (un)observable subspaces. In contrast, in this
paper we propose constructive algebraic conditions for the synthesis of coherent
feedback to equip the system with a DFS. These conditions are expressed in
terms of linear matrix inequalities (LMIs) and reduce the DFS synthesis problem
to an algebraic pole assignment problem. After completing this work we
became aware that Nurdin and Gough had also arrived at the pole placement
idea~\cite{Nurdin2015}. However, our results are different in that they are
not restricted to interconnected optical cavities considered
in~\cite{Nurdin2015}, and applicable to a coherent feedback interconnection of
two general quantum systems of which interconnected optical
cavities are a special case; see Section~\ref{sec:ex}. Of course, the
generality of our formulation means that the DFS engineering problem in
this paper cannot be solved by calculating system poles directly,
hence a more general approach is developed in this paper.

Also, the DFS synthesis methodology proposed here extends
substantially our preliminary work~\cite{Pan2016}. The controller
configuration in that paper was limited to resembling a classical
Luenberger observer. It turns out that such a configuration is somewhat
restrictive; for example, it is not sufficiently flexible  to
capture the controller structure analyzed in~\cite{Yamamoto2014}. In this paper, we
build our technique using the most general type of dynamic linear passive
coherent feedback. We show that the controller structures
from~\cite{Pan2016,Yamamoto2014} are in fact special cases of our general 
setting. In addition, we discuss the conventional open-loop approach to
reservoir engineering and show the shortcoming of such approach. A
shortened version of this paper has been scheduled for presentation at the
2017 American Control Conference~\cite{Nguyen2017(toappear)}. Compared to the
conference version, the present version is substantially revised and
extended. In particular, it includes background material on quantum passive
systems and complete proofs of results. Also, a new example is included to
illustrate the possibility of creating a DFS shared by the principal plant
and the controller, which appears to be not possible to achieve in simple
optical cavity systems.

{\em Notation}. Given an underlying Hilbert space $\mathfrak{H}$ and an
operator $x\colon \mathfrak{H}\to \mathfrak{H}$, $x^*$ denotes the
operator adjoint to $x$. In the case of a vector of operators, the vector
consisting of the adjoint 
components of $x$ is denoted $x^\#$, and $x^\dagger=(x^\#)^T$, where $^T$
denotes the transpose of a vector. Likewise, for a matrix $A$, $A^\#$ is
the matrix whose entries are complex conjugate of the corresponding entries
of $A$, and $A^\dagger=(A^\#)^T$. $[x,y]=xy-yx$ is the commutator of two
operators, and in the case where $x, y$ are vectors of operators,
$[x,y^\dagger]=xy^\dagger -(y^\#x^T)^T$.

\section{Background}

\subsection{Open Quantum Systems}
\label{sec:open}

Open quantum systems are systems that are coupled to an external
environment or reservoir \cite{Breuer2002}. The environment exerts an influence
on the system, in the form of vectors $W(t)$, $W^\dagger(t)$  consisting of
quantum Wiener processes defined on a Hilbert space $\mathfrak{F}$ 
known as the Fock space. The unitary motion of the passive annihilation
only system governed by these processes is described by the stochastic
differential equation   
\begin{eqnarray}
  \label{eq:U}
 dU(t)\! &=&\! \left(\!(-iH\! -\frac{1}{2}L^\dagger L)dt \! + \!
   dW^\dagger L\! -\! L^\dagger dW \! \right)\! U(t), \quad \\
 U(0)&=& I, \nonumber 
\end{eqnarray}
where $H$ and $L$ are, respectively, the system Hamiltonian and the
coupling operator through which the system couples to the
environment. Then, any operator $X\colon \mathfrak{H}\to \mathfrak{H}$
generates the evolution 
$X(t)=j_t(X)=U(t)^*(X\otimes I)U(t)$ in the space of operators on the tensor
product Hilbert space $\mathfrak{H}\otimes \mathfrak{F}$,
\begin{align}
d X = \mathcal{G}(X)dt+ dW^\dagger [X,L] + [L^\dagger, X] dW, \label{Heis.X}
\end{align}
where 
\begin{eqnarray*}
 \mathcal{G}(X)&=&-i [X,H] + \mathscr{L}_L(X), \\
 \mathscr{L}_L(X) & =& \frac{1}{2} L^\dagger [X,L] + \frac{1}{2} [L^\dagger, X]L
\end{eqnarray*}
are the generator and the Lindblad superoperator of the system,
respectively~\cite{Wiseman2009}.  The field resulting from the 
interaction between the system and the environment constitutes the output
field of the system  
\begin{eqnarray}
dY= L dt + dW. 
\end{eqnarray}

\subsection{Linear annihilation only systems}

Linear annihilation only systems arise as a particular class of open
quantum systems whose operators $a_k$, $k=1,\ldots, n$, describe various modes
of photon annihilation resulting from interactions between the
environment and the system. Such operators satisfy the canonical
commutation relations  
$[a_j, a_k^*]=\delta_{jk}$, where $\delta_{jk}$  is the Kronecker
delta. Taking the system Hamiltonian and the coupling operator of the
system to be, respectively, quadratic and linear functions of the vector
$X=a=\left[a_{1},\ldots a_{n}\right]^{T}$, 
\begin{equation}
   \label{eq:HL}
   H=a^\dagger Ma, \quad L= Ca, 
\end{equation}
where $M$ is a Hermitian $n\times n$ matrix, and $C\in \mathbb{C}^{m\times
  n}$, the dynamics and output equations become  
\begin{eqnarray} 
da &=&  A adt + B dW \nonumber \\
dy &=& Ca dt + dW,
\label{eq:passive-sys}
\end{eqnarray}
where the complex matrices $A\in \mathbb{C}^{n\times n}$, $B\in
\mathbb{C}^{n\times m}$, and $C\in \mathbb{C}^{m\times n}$ satisfy 
\begin{eqnarray}
A= -i M  - \frac{1}{2}  C^\dagger  C, \ \ B = -C^\dagger .
\label{pr.cond}
\end{eqnarray}
The following fundamental identity then holds~\cite{Maalouf2011}
\begin{eqnarray}
 A+ A^\dagger + C^\dagger C =0.
\label{eq:pr-passive}
\end{eqnarray}

\subsection{Passive annihilation only quantum systems}

According to~\cite{James2010}, passivity of a quantum system $\mathbf{P}$ is 
defined as a property of the system with respect to an output generated by
an exosystem $\mathbf{W}$ and applied to input channels of the given
quantum system on one hand, and a performance operator $Z$ of the system on
the other hand. To 
particularize the definition of~\cite{James2010} in relation to the specific
class of annihilation only systems, we consider a class of
exosystems, i.e., open quantum systems with zero Hamiltonian,
an identity scattering matrix and a coupling operator $u$ which couples the
exosystem with its input field. The exosystem is assumed to be independent
of $\mathbf{P}$ in the sense that $u$ commutes with any 
operator from the $C^*$ operator algebra generated by $X$ and $X^\dagger$.    
The time evolution of $u$ is however determined by the full
interacting system $\mathbf{P}\triangleleft \mathbf{W}$, and therefore may be influenced by
$X, X^\dagger$. 

If the output of the exosystem $\mathbf{W}$ is fed into the input of
the system $\mathbf{P}$ in a cascade
or series connection, the resulting system 
$\mathbf{P}\triangleleft \mathbf{W}$ has the  Hamiltonian $H_{\mathbf{P}\triangleleft
  \mathbf{W}}=H+\mathrm{Im}(u^\dagger L)$, the identity scattering matrix and the field coupling operator
$L_{\mathbf{P}\triangleleft  \mathbf{W}}=L+u$~\cite{James2010}. The resulting system ($\mathbf{P}\triangleleft \mathbf{W}$) then has the generator $\mathcal{G}_{\mathbf{P}\triangleleft \mathbf{W}}$.

\begin{Def}[\cite{James2010}]\label{passivity.def}
  A system $\mathbf{P}$ with a performance output $Z$ is passive if there exists
  a nonnegative observable $V$ (called the \emph{storage observable} of $P$)
  such that  
  \begin{equation}
    \label{passivity.cond}
    \mathcal{G}_{\mathbf{P}\triangleleft \mathbf{W}} (V)\le  Z^\dagger u +u^\dagger
    Z+\lambda 
  \end{equation}
for some constant $\lambda>0$. The operator 
\[
r(\mathbf{W})=Z^\dagger u +u^\dagger Z 
\]
is the \emph{supply rate} which ensures passivity.  
\end{Def}

Now suppose $\mathbf{P}$ is a linear annihilation only system
(\ref{eq:HL}). Also, consider a performance output for the system
$\mathbf{P}\triangleleft \mathbf{W}$  to be
\begin{equation*} 
Z=C_0a+D_0u,
\end{equation*}
with $C_0\in \mathbb{C}^{l\times n}$, $D_0\in \mathbb{C}^{l\times m}$. 
Taking $X=a$ in (\ref{Heis.X}), the system
$\mathbf{P}\triangleleft \mathbf{W}$ can be written as
\begin{eqnarray} 
d a &=&  (A a + B u)dt + BdW, 
 \label{eq:passive-sys.1} \\
 dY &=& (C a + u)dt + dW, \nonumber \\
Z&=&C_0a+D_0u.\nonumber
\end{eqnarray}
where the complex matrices $A\in \mathbb{C}^{n\times n}$, $B\in
\mathbb{C}^{n\times m}$, and $C\in \mathbb{C}^{m\times n}$ are the
coefficients of the annihilation only system $\mathbf{P}$.

We further take the storage observable $V$ having the form $V=a^\dagger Pa$, and the supply rate having the form $r(\mathbf{W})=Z^\dagger u
+u^\dagger Z$. Then it can be shown that the system $\mathbf{P}$ is passive
with a storage function $V$ and a supply rate $r(\mathbf{W})$ 
if for some constant $\lambda>0$, 
\begin{eqnarray*}
\lefteqn{a^\dagger (PA+A^\dagger P) a + u^\dagger BPa+a^\dagger PBu} && \\
&&\le (C_0a+D_0u)^\dagger u + u^\dagger (C_0a+D_0u)+\lambda.
\end{eqnarray*}
This condition is equivalent to the positive realness
condition stated in Theorem 3 of~\cite{Zhang2011} (letting $Q=0$ in that theorem): 
\begin{eqnarray}
\left[
  \begin{array}{cc}
    PA+A^\dagger P & PB-C_0^\dagger \\
    B^\dagger P-C_0 & -(D_0+D_0^\dagger) \end{array}
  \right]\le 0.
\end{eqnarray}
In the special case, where $V=a^\dagger a$, 
$D_0=0$~\cite{Zhang2011} and $C_0=-C$, this reduces to the following inequality
\[
A+A^\dagger\le 0
\]
as the condition for passivity. Clearly this condition is satisfied in the
case of an annihilation only system $\mathbf{P}$ in the light of the
identity (\ref{eq:pr-passive}). 
Hence the annihilation only system (\ref{eq:passive-sys.1}) is passive with
respect to performance output $Z=-Ca$, with the storage function
$V=a^\dagger a$.

\subsection{Decoherence free subsystems}
\label{sec:dfs}

As mentioned, a decoherence free subsystem represents a subsystem whose
variables are not affected by input fields and do not appear in the system
output fields; this makes the DFS isolated from the environment and
inaccessible to measurement devices, thus preserving the quantum
information carried by the variables of the DFS. In relation to the
annihilation only system (\ref{eq:passive-sys}), with
$a=[a_1,\ldots,a_n]^T$, a component $a_j$ is a decoherence-free mode if the
evolution of $a_j$ is independent of the input  $W$ and if the system
output $Y$ is independent of $a_j$. The collection of decoherence-free
modes forms a subspace, called the {\em decoherence-free
  subspace}. 

An important fact about the existence of a decoherence-free
subsystem for linear annihilation only systems follows from the results
established in~\cite{Gough2015}:

\begin{Pro}\label{Gough2015-prop}
The linear annihilation only system (\ref{eq:passive-sys}) has a
decoherence-free subsystem if and only if the matrix $A$ has some of its
poles on the imaginary axis, with the remaining poles residing in the open
left half-plane of the 
complex plane. 
\end{Pro}

\emph{Proof: }  
According to~\cite[Lemma~2]{Gough2015}, for the system (\ref{eq:passive-sys}),
the properties of controllability, observability and Hurwitz stability are
equivalent. The statement of the proposition then follows by
contraposition, after noting that being passive, the system
(\ref{eq:passive-sys}) cannot have eigenvalues in the open right hand-side
of the complex plane due to (\ref{eq:pr-passive}).   
\hfill$\Box$

According to Proposition~\ref{Gough2015-prop}, if the system 
(\ref{eq:passive-sys}) has a DFS, then there must exist a coordinate
transformation of the system (\ref{eq:passive-sys}) such that in the new
coordinates, the system takes the form, known as the Kalman decomposition:
\begin{eqnarray*} 
d\tilde a &=&  \left[\begin{array}{cc}\tilde A_{11} & \tilde A_{12} \\
0 & \tilde A_{22}
\end{array}\right]
\tilde adt + \left[\begin{array}{c}\tilde B_1 \\ 0
  \end{array}\right]
  dW \nonumber \\
dy &=& \left[\begin{array}{cc}\tilde C_1 & 0
  \end{array}\right]
  \tilde a dt + dW,
\label{eq:passive-sys.KD}
\end{eqnarray*} 
By partitioning the vector $\tilde a$ accordingly, $\tilde a=[\tilde
a_1~\tilde a_2]$, we observe that  the decoherence induced by the environment
and probing fields will not affect dynamics of the operator $\tilde a_2$.   
Furthermore, by expressing the system Hamiltonian $H$ in the new
coordinates as $H=\tilde a^\dagger \tilde M\tilde a$, we observe   
from the corresponding equation (\ref{pr.cond}) that  
\[
\tilde A_{22}=-i \tilde M_{22},
\]
where $\tilde M_{22}$ is the corresponding block of the matching partition
of the matrix $\tilde M$. Since $\tilde M_{22}$ is Hermitian and has only
real eigenvalues, this implies that the matrix $\tilde A_{22}$ can only
have imaginary eigenvalues. This observation suggests that engineering a
quantum system to have a decoherence free amounts to placing some of the
poles of the corresponding system (\ref{eq:passive-sys}) on the
imaginary axis.

\section{Coherent reservoir engineering}

Reservoir engineering refers to the process of determining and implementing
coupling operators $L = [L_1;...;L_n]$ for an open quantum system such that
desired behaviors are achieved. Examples of common objectives include quantum computation by dissipation \cite{Verstraete2009}, entanglement \cite{Krauter2011}, state preparation \cite{Ticozzi2009}, and protection of quantum information \cite{Cohen2014, Pan2016a}. Typically open systems have some unavoidable couplings to the environment, and such channels may lead to loss of energy and quantum coherences. However, in many systems couplings can be engineered at the fabrication stage, providing a resource for tuning the behavior of the system.

\begin{figure}[t]
\begin{center}
\psfrag{S}{$S$}
\psfrag{W}{$W$}
\psfrag{Quantum}{\hspace{-2ex}Quantum}
\psfrag{system}{\hspace{-2ex}system}
\psfrag{controller}{\hspace{-2ex}controller}
\psfrag{w}{$w$}
\psfrag{u}{$u$}
\psfrag{y}{$y$}
\psfrag{f}{$f$}
\psfrag{s}{$s$}
\psfrag{r}{$r$}
\psfrag{z}{$z$}
\psfrag{v}{$v$}
\psfrag{y'}{$y'$}
\psfrag{z'}{$z'$}
\psfrag{u'}{$u'$}
\psfrag{u1'}{$\tilde u'$}
\psfrag{u1}{$\tilde u$}
\includegraphics[width=0.5\columnwidth]{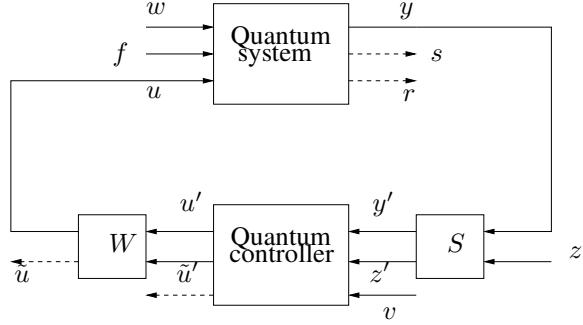}
\caption{Coherent feedback network for DFS generation.}
\label{fig:obs-dfs1}
\end{center}
\end{figure}

In this section, the main results of the paper are presented. With
reference to Fig.~\ref{fig:obs-dfs1}, we investigate conditions to enable
the synthesis of a quantum coherent controller-system network to generate a
DFS in the interconnected system through interactions between the principal
quantum system and the controller.

The quantum linear passive system in Fig.~\ref{fig:obs-dfs1} is the system
of the form (\ref{eq:passive-sys}), and its input fields are
further partitioned as $W=[w^T,u^T,f^T]^T$. Here, $w$ represents a
`natural' environment for the system, and $f$ and $u$ represent an
open-loop and feedback engineered fields, respectively. According to this
partitioning,  the system evolution is described as
\begin{subequations}
\label{eq:annihilationplant}
\begin{align}
&d{a}_p=A_pa_pdt+B_1dw + B_2 du +B_3df,\\ 
&dy=C_pa_pdt+dw.
\end{align}
\end{subequations}
Accordingly, the matrices of the system have dimensions as follows: $A_p\in
\mathbb{C}^{n\times n}$, $B_1\in 
\mathbb{C}^{n\times n_w}$, $B_2\in \mathbb{C}^{n\times n_u}$ $B_3\in
\mathbb{C}^{n\times n_f}$, and $C_p\in \mathbb{C}^{n_w\times n}$ ($n, n_w,
n_u, n_f \in \mathbb{N}$). We also use the notation $a_p$ for the vector 
$a_p\left(t\right)=\left[a_{p_1}\left(t\right),\ldots
  a_{p_n}\left(t\right)\right]^{T}$ of the system annihilation operators
defined on its underlying Hilbert space $\mathfrak{H}_p$. 

In terms of the Hamiltonian and coupling operators, the system has the
Hamiltonian 
\begin{equation}
H_p = a_p^{\dagger}Ma_p
\label{ham.p}
\end{equation}
where $M$ is an $n\times n$ complex
Hermitian matrix, and is linearly coupled to the input fields via the
coupling operators 
\begin{equation}
L_{p_1}=\alpha_1 a_p, \quad L_{p_2}=\alpha_2 a_p, \quad L_{p_3}=\alpha_3 a_p
\label{capl.p}
\end{equation}
where $\alpha_1 \in \mathbb{C}^{n_w \times n}$, $\alpha_2 \in
\mathbb{C}^{n_u \times n}$, $\alpha_3 \in \mathbb{C}^{n_f \times n}$ are
complex matrices. Then the relations (\ref{pr.cond}) specialize as follows: 
\begin{align}
A_p&=-\left(iM+\frac{1}{2}\alpha_1^{\dagger}\alpha_1 +
  \frac{1}{2}\alpha_2^{\dagger}\alpha_2+
  \frac{1}{2}\alpha_3^{\dagger}\alpha_3\right),\nonumber\\ 
B_1&=-\alpha_1^{\dagger},\nonumber\\
B_2&=-\alpha_2^{\dagger},\nonumber\\
B_3&=-\alpha_3^{\dagger},\nonumber\\
C_p&=\alpha_1.\nonumber
\end{align}

The starting point of the discussion that follows is the assumption that
under the influence of its natural environment $w$ alone, (i.e., in the
absence of the engineered fields $f$ and $u$), the system does not possess
a DFS. Mathematically, this assumption corresponds to the assumption that    
$\left(A_p,B_1\right)$ is controllable and $\left(A_p,C_p\right)$ is observable,
since these properties rule out the existence of a DFS in the
plant~(\ref{eq:annihilationplant}) when $B_2=0$, $B_3=0$; see
Proposition~\ref{Gough2015-prop} and \cite{Yamamoto2014,Gough2015}.

\subsection{Open loop reservoir engineering for DFS generation}
\begin{figure}[t]
\begin{center}
\psfrag{Quantum}{\hspace{-2ex}Quantum}
\psfrag{system}{\hspace{-2ex}system}
\psfrag{w}{$w$}
\psfrag{y}{$y$}
\psfrag{f}{$f$}
\psfrag{s}{$s$}
\includegraphics[width=0.35\columnwidth]{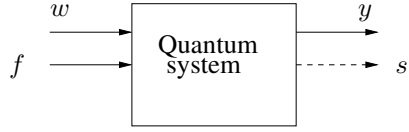}
\caption{Open loop setup for DFS generation.}\label{fig:static-dfs}
\end{center}
\end{figure}

In many cases, system couplings can be engineered at a fabrication stage to
reduce unavoidable loss of energy due to decoherence~\cite{Poyatos1996,Verstraete2009}. 
The process of tuning the system at the fabrication stage does not involve
feedback, and we let $L_{p_2}=0$, which corresponds to $\alpha_2=0$ and $B_2=0$
in (\ref{eq:annihilationplant}); see Fig.~\ref{fig:static-dfs}.
 
Then the
system (\ref{eq:annihilationplant}) reduces to that of the form 
\begin{subequations}
\label{eq:annihilationplant.static}
\begin{align}
&d{a}_p=A_pa_pdt+B_1dw+B_3df,\\ 
&dy=C_pa_pdt+dw
\end{align}
\end{subequations}
Here, $w$ and $f$ symbolize the natural environment and the
fabricated open-loop field, respectively. Accordingly, the coupling
operator $L_{p_1}$ corresponds to a fixed coupling with the natural
environment, while the coupling $L_{p_3}$ corresponds to the engineered
coupling. The physical realizability requirement 
imposes the constraint that 
\begin{equation} 
A_p+A_p^\dagger+B_1B_1^\dagger+B_3B_3^\dagger=0,
\label{prc.static}
\end{equation}
cf.~(\ref{eq:pr-passive}). Recall~\cite{Maalouf2011} that a quantum stochastic
differential equation of the form (\ref{eq:annihilationplant.static}) is
said to be (canonically) physically realizable if it preserves the canonical
commutation relations,  
$
[a_p,a_p^\dagger]=a_pa_p^\dagger-(a_p^* a_p^T)^T=I, 
$    
and is a representation of an open harmonic oscillator, i.e., it possesses
a Hamiltonian and a coupling operator. The satisfaction of the identity
(\ref{prc.static}) is a necessary and sufficient condition
for physical realizability~\cite[Theorem~5.1]{Maalouf2011}.

\begin{Thm}\label{T1} 
Suppose $(-iM,B_1)$ is controllable. Then 
a DFS cannot be created by coupling the system to an engineered
environment.
\end{Thm}

\emph{Proof: }
To prove the theorem we will show that the matrix $A_p$ has all its
eigenvalues in the open left half-plane of the complex plane, and therefore
it cannot have a DFS, according to Proposition~\ref{Gough2015-prop};
see~\cite[Lemma~2]{Gough2015}.  
  
First consider the system with a fixed coupling with
the environment, i.e.,  $L_{p_3}=0$. For this system, the physical
realizability properties dictate that 
\begin{equation} 
A_{p1}+A_{p1}^\dagger+B_1B_1^\dagger=0,  
\label{prc.static.1}
\end{equation}
with $A_{p1}=-iM-\frac{1}{2}B_1B_1^\dagger$; see (\ref{pr.cond}). 

Recall that for an arbitrary $n\times n$ matrix $\Phi$ and an $n\times m$
matrix $B$, the pair $(\Phi,B)$ is controllable if and only if
$(\Phi+\frac{1}{2}BB^\dagger,B)$ is controllable.  
Applying this fact to the pair $(-iM,B_1)$ which is controllable by the
assumption of the theorem, we conclude that 
$(A_{p1},B_1)$ is controllable. Thus, equation
(\ref{prc.static.1}) can be regarded as a Lyapunov equation
\[
A_{p1}P+PA_{p1}^\dagger+B_1B_1^\dagger=0
\]
with controllable $(A_{p1},B_1)$, which has a positive definite solution $P=I$.
Since  $B_1B_1^\dagger\ge 0$, according to the inertia
theorem~\cite[Theorem 3]{Chen1973}, the above observation about the
existence of a positive definite solution to the Lyapunov equation implies
that $A_{p1}$ must have all its eigenvalues in the open 
left half-plane of the complex plane. As a result, if $(-iM,B_1)$ is
controllable, the corresponding passive quantum system 
with fixed coupling cannot have a DFS, according to
Proposition~\ref{Gough2015-prop}. 

Next consider this system when it is coupled to an engineered environment,
i.e., $L_{p_3}\neq 0$ and $B_3\neq 0$. Since $A_{p1}$ has been shown to have
all eigenvalues in the open left half-plane of the complex plane, there
exists a positive definite Hermitian matrix $P=P^\dagger>0$ such that 
\[
A_{p1}^\dagger P+PA_{p1}<0.
\]  
On the other hand, according to Corollary~4 of~\cite{Ostrowski1962}, the matrix
$\frac{1}{2}B_3B_3^\dagger P$ cannot have eigenvalues in the open left
half-plane of the complex plane, and therefore $-\frac{1}{2}B_3B_3^\dagger
P-\frac{1}{2}PB_3B_3^\dagger\le 0$. This implies that
\[
(A_{p1}-\frac{1}{2}B_3B_3^\dagger)^\dagger P+P(A_{p1} -\frac{1}{2}B_3B_3^\dagger)
<0
\]    
and therefore $A_p=A_{p1} -\frac{1}{2}B_3B_3^\dagger$ must have all its
eigenvalues in the open left half-plane of the complex plane. According to
Proposition~\ref{Gough2015-prop}, this rules out the possibility for the system
with engineered coupling to have a DFS. 

\hfill$\Box$

Next, suppose that $(-iM,[B_1 ~B_3])$ is not controllable\footnote{Here,
  $[B_1 ~B_3]$ is the matrix obtained by concatenating the rows of $B_1$
  and $B_3$.}, therefore 
$(-iM,B_1)$ is not controllable either. Theorem~\ref{T1} does not rule out
a possibility for a DFS to exist in this case. It is easy to show that 
\[
\mathrm{ker}(\mathcal{C}^T)= 
\mathrm{ker}(\mathcal{C}_w^T)\cap \mathrm{ker}(\mathcal{C}_f^T),
\]
where $\mathcal{C}_w$, $\mathcal{C}_f$ are the controllability matrices
with respect to the inputs $w$ and $f$, respectively. 
From this observation, it follows that the dimension of the DFS of system
(\ref{eq:annihilationplant.static}) is less or equal to the dimension of
each of the decoherence free subsystems arising when the quantum plant is
coupled with the fixed and engineered fields only. This leads to the
conclusion that coupling the system with additional engineered fields can
only reduce the dimension of the DFS. In the remainder of the paper, we
will show that using coherent feedback, on the other hand, does allow to
create or increase dimension of a DFS.

\subsection{Coherent feedback reservoir engineering}

In this section we consider a system of the form
(\ref{eq:annihilationplant}). To simplify the notation we will combine two
static channels $w$ and $f$ into a single channel, which will again be
denoted as $w$. More precisely, we combine the coupling operators $L_{p_1}$
and $L_{p_3}$ into a single operator $L_{p_1}$. Then the system
(\ref{eq:annihilationplant}) reduces to a system of the form
\begin{subequations}
\label{eq:annihilationplant.dyna}
\begin{align}
&d{a}_p=A_pa_pdt+B_1dw + B_2 du,\\ 
&dy=C_pa_pdt+dw,
\end{align}
\end{subequations}
where the new matrix $B_1$ is composed of the previous matrices $B_1$ and
$B_3$, so that using the new notation we have 
\begin{align}
A_p&=-\left(iM+\frac{1}{2}\alpha_1^{\dagger}\alpha_1 +
  \frac{1}{2}\alpha_2^{\dagger}\alpha_2\right),\nonumber\\ 
B_1&=-\alpha_1^{\dagger},\nonumber\\
B_2&=-\alpha_2^{\dagger},\nonumber\\
C_p&=\alpha_1.\label{plant.prc}
\end{align}

For a coherent quantum controller for the quantum plant
\eqref{eq:annihilationplant}, we will consider another open quantum linear
annihilation only system. Such a system will be assumed to be coupled with three
environment noise channels, $y'$, $z'$ and $v$. The fields $y'$, $z'$ are to
produce output fields which will be used to form the feedback, and the
channel $v$ will be used to ensure that the constructed
observer is physically realizable. As is known~\cite{Maalouf2011}, once physical
realizability of the observer is ensured, one can readily construct a
scattering matrix, a Hamiltonian and a collection of coupling operators
describing the quantum evolution of the controller in the form of a quantum
stochastic differential equation (\ref{Heis.X}). Alternatively, a
physically realizable coherent controller can be represented in the form of
the quantum stochastic differential equation
(\ref{eq:passive-sys})~\cite{Maalouf2011}, i.e., in the form           
\begin{subequations}
\label{eq:annihilationobserver}
\begin{align}
d{a}_c=&A_ca_cdt+G_1dy'+G_2dz' + G_3dv,\\ 
du'=&Ka_cdt+dy', \\
d\tilde u'=&\tilde Ka_cdt+dz',
\end{align}
\end{subequations}
where for physical realizability, the following constraints must be
satisfied~\cite[Theorem~5.1]{Maalouf2011}:  
\begin{eqnarray}
&&  A_c+A_c^\dagger+G_1G_1^\dagger + G_2G_2^\dagger+ G_3G_3^\dagger
=0, \label{contr.prc} \\
&&K=-G_1^\dagger, \\
&&\tilde K=-G_2^\dagger.
\end{eqnarray}

Interconnection between the controller and the plant are through scattering
equations relating the output fields of the plant with the input channels
of the controller and \emph{vice versa}. Specifically, the scattering
equation  
\begin{eqnarray}
  \left[
    \begin{array}{c}
      y' \\ z'
    \end{array}
\right]= S  \left[
    \begin{array}{c}
      y \\ z
    \end{array}
\right],
\end{eqnarray}
links the output field of the plant $y$ and the controller environment
$z$ with the input controller channels $y'$, $z'$. Here, $S$ is a unitary
matrix partitioned as 
\begin{eqnarray}
  \label{eq:S}
  S=  \left[    \begin{array}{cc}
S_{11} & S_{12} \\ S_{21} & S_{22}
    \end{array}
\right].
\end{eqnarray}
Likewise, feedback from the controller (\ref{eq:annihilationobserver}) is
via a unitary matrix  $W$, 
\begin{eqnarray}
  \left[
    \begin{array}{c}
      u \\ \tilde u
    \end{array}
\right]= W  \left[
    \begin{array}{c}
      u' \\ \tilde u'
    \end{array}
\right], \quad   W=  \left[    \begin{array}{cc}
W_{11} & W_{12} \\ W_{21} & W_{22}
    \end{array}
\right].
\end{eqnarray}

The matrices $A_c$, $G_1=-K^\dagger$, $G_2=-\tilde K^\dagger$, $G_3$,
and the scattering matrices $S$, $W$ are regarded as the controller design
parameters. Our objective in this paper is to find a procedure for
selecting those parameters so that the resulting coherently interconnected
quantum system in Fig.~\ref{fig:obs-dfs1} possesses a decoherence free
subsystem.

To devise the DFS synthesis procedure, we first note that the control
system governed by $y,z,v$ and output $u$ can be 
represented as
 \begin{eqnarray}
\label{eq:annihilationobserver.yz}
d{a}_c&=&(A_ca_c-(G_1S_{11}+G_2S_{21})B_1^\dagger a_p)dt
\nonumber \\
&& +(G_1S_{11}+G_2S_{21})dw\nonumber \\
&& +(G_1S_{12}+G_2S_{22})dz+ G_3dv,\\
du&=&(-(W_{11}G_1^\dagger+W_{12}G_2^\dagger)a_c \nonumber \\
&& -(W_{11}S_{11}+W_{12}S_{21})B_1^\dagger a_p)dt \nonumber \\
&& 
+(W_{11}S_{11}+W_{12}S_{21})dw\nonumber \\
&& +(W_{11}S_{12}+W_{12}S_{22})dz\nonumber
\end{eqnarray}
Also, the closed loop system is described by the quantum stochastic
differential equation 
 \begin{eqnarray}
\label{closed.loop}
d\left[ \begin{array}{c}  a_p \\
 a_c     \end{array}
\right]
= A_{cl}\left[ \begin{array}{c}  a_p \\
 a_c     \end{array}
\right] dt
+B_{cl} \left[ \begin{array}{c}  dw \\
 dz \\ dv     \end{array}  
\right], 
\end{eqnarray}
with block matrices $A_{cl}$, $B_{cl}$ partitioned as shown
in~(\ref{cl.partition}):

 \begin{eqnarray}
A_{cl}&=&
\left[ \begin{array}{c|c}
{\begin{aligned}
\quad A_p -B_2(W_{11}S_{11}+W_{12}S_{21})B_1^\dagger \quad
\end{aligned}}
&
{\begin{aligned}
\quad  -B_2(W_{11}G_1^\dagger+W_{12}G_2^\dagger) \quad 
\end{aligned}}  \\ \hline 
{\begin{aligned}
-(G_1S_{11}+G_2S_{21})B_1^\dagger
\end{aligned}} 
&
{\begin{aligned}
A_c 
  \end{aligned}}
\end{array}
\right], \nonumber \\[10pt]
B_{cl}&=&
\left[ \begin{array}{c|c|c}
{\begin{aligned}
\quad B_1+B_2(W_{11}S_{11}+W_{12}S_{21}) \quad
\end{aligned}}
&
{\begin{aligned}
\quad B_2(W_{11}S_{12}+W_{12}S_{22}) \quad
\end{aligned}}  
&
{\begin{aligned}
\quad 0 \quad
\end{aligned}}
\\ \hline 
{\begin{aligned}
\quad G_1S_{11}+G_2S_{21} \quad
\end{aligned}} 
&
{\begin{aligned}
\quad G_1S_{12}+G_2S_{22} \quad
  \end{aligned}}
&
{\begin{aligned}
\quad G_3 \quad
  \end{aligned}}
\end{array}
\right].
\label{cl.partition}
\end{eqnarray}

\begin{Lem}
\label{hat-and-check}
Let
\begin{eqnarray}
  \label{Ao}
  A_c=A_p &-& B_2(W_{11}S_{11}+W_{12}S_{21})B_1^\dagger \nonumber  \\
&& +(G_1S_{11}+G_2S_{21})B_1^\dagger \nonumber \\
&& -B_2(W_{11}G_1^\dagger+W_{12}G_2^\dagger) 
\end{eqnarray}
Then for $A_{cl}$ to have all eigenvalues on the imaginary axis or in the
left half-plane of the complex plane it is
necessary and sufficient that the following matrices 
\begin{eqnarray}
\hat A&=&A_p-B_2(W_{11}S_{11}+W_{12}S_{21})B_1^\dagger \nonumber \\ 
&& -B_2(W_{11}G_1^\dagger + W_{12}G_2^\dagger),   \label{eq:Ahat} \\
\check A&=&A_p-B_2(W_{11}S_{11}+W_{12}S_{21})B_1^\dagger \nonumber  \\
&& +(G_1S_{11}+G_2S_{21})B_1^\dagger \label{eq:Acheck}
\end{eqnarray}
have all eigenvalues on the imaginary axis or in the left half-plane of the
complex plane.  
\end{Lem}

\emph{Proof: }
The matrix $A_{cl}$ has the same eigenvalues as the matrix 
\begin{eqnarray*}
\lefteqn{\left[ \begin{array}{cc}I & 0 \\ I & -I \end{array} \right]
A_{cl}
\left[ \begin{array}{cc}I & 0 \\ I & -I \end{array} \right]} && \\
&&=\left[ \begin{array}{c|c}
{\scriptsize \begin{aligned}
A_p-B_2W_{11}(S_{11}B_1^\dagger+G_1^\dagger)\\ 
-B_2W_{12}(S_{21}B_1^\dagger+G_2^\dagger)
\end{aligned}}
&
{\scriptsize \begin{aligned}B_2(W_{11}G_1^\dagger+W_{12}G_2^\dagger)
  \end{aligned}}  \\[10pt] \hline 
{\scriptsize\begin{aligned}
\phantom{^{T^{T^T}}}A_p-A_c\\
-B_2(W_{11}S_{11}+W_{12}S_{21})B_1^\dagger \\
+(G_1S_{11}+G_2S_{21})B_1^\dagger\\
-B_2(W_{11}G_1^\dagger+W_{12}G_2^\dagger)
\end{aligned}} 
&
{\scriptsize \begin{aligned}B_2(W_{11}G_1^\dagger+W_{12}G_2^\dagger)+A_c
  \end{aligned}}
\end{array}
\right].
\end{eqnarray*}
Hence the lemma follows, due to the definition of $A_c$ in~(\ref{Ao}).

\hfill $\Box$

\begin{Thm}\label{DFS.synthesis.theorem}
Suppose matrices $S$, $W$ are given. Let $G_1$, $G_2$ be such that
\begin{enumerate}[(a)]
\item
The following linear matrix inequality (LMI) in $G_1$, $G_2$ is satisfied
\begin{eqnarray}
  \label{LMI}
\left[\begin{array}{ccc}
R & G_1 & G_2\\
G_1^\dagger & -I & 0 \\
G_2^\dagger & 0 & -I
\end{array}
\right]\le 0,
\end{eqnarray}
where
\begin{eqnarray}
\lefteqn{R=
-B_1B_1^\dagger -B_2B_2^\dagger - B_2(W_{11}S_{11}+W_{12}S_{21})B_1^\dagger} && \nonumber \\
&& 
- B_1(S_{21}^\dagger W_{12}^\dagger+S_{11}^\dagger W_{11}^\dagger)B_2^\dagger
+(G_1S_{11}+G_2S_{21})B_1^\dagger \nonumber \\ 
&&
+ B_1(S_{11}^\dagger G_1^\dagger+S_{21}^\dagger G_2^\dagger) 
-B_2(W_{11}G_1^\dagger+W_{12}G_2^\dagger) \nonumber \\ 
&&
-(G_1W_{11}^\dagger + G_2W_{12}^\dagger)B_2^\dagger;
\end{eqnarray}

\item
The matrices $\hat A$ and $\check A$, defined in equations~(\ref{eq:Ahat})
and~(\ref{eq:Acheck}) respectively, have all their eigenvalues in the 
closed left half-plane, with at least one of them having  eigenvalues on
the imaginary axis.  
\end{enumerate}
Then a matrix $G_3$ can be found such that the closed loop system
(\ref{closed.loop}) admits a DFS.
\end{Thm}

\emph{Proof: }
Via the Schur complement, (\ref{LMI}) is equivalent to 
\[
R+G_1G_1^\dagger+G_2G_2^\dagger\le 0.
\]  
Therefore one can find $G_3$ such that 
\[
R+G_1G_1^\dagger+G_2G_2^\dagger+G_3G_3^\dagger= 0.
\]  
From this identity and the expression (\ref{Ao}), the identity
(\ref{contr.prc}) follows. This shows that the feasibility of the LMI
(\ref{LMI}) ensures that the controller system
(\ref{eq:annihilationobserver}) can be made physically realizable by
appropriately choosing $G_3$. As a result, the closed loop system, being a
feedback interconnection of physically realizable systems, is a
physically realizable annihilation only system. Also, condition (b) and
Lemma~\ref{hat-and-check} ensure that $A_{cl}$ has eigenvalues on the
imaginary axis. Then it follows from Proposition~\ref{Gough2015-prop} that the closed
loop system (\ref{closed.loop}) has a DFS.

\hfill$\Box$

Note that matrices $\hat A$ and $\check A$ can be rewritten as
\begin{eqnarray}
\hat A&=&A_p-B_2(W_{11}S_{11}+W_{12}S_{21})B_1^\dagger \nonumber \\ 
&& -B_2[W_{11}~W_{12}]\left[\begin{array}{c}G_1^\dagger \\ G_2^\dagger
    \end{array}\right],
    \label{eq:Ahat.contr} \\
\check A&=&A_p-B_2(W_{11}S_{11}+W_{12}S_{21})B_1^\dagger \nonumber  \\
&& +[G_1~G_2]\left[\begin{array}{c} S_{11} \\ S_{21}\end{array}\right]
B_1^\dagger \label{eq:Acheck.obsrv}
\end{eqnarray}
A necessary condition to ensure that an eigenvalue assignment can be carried out
for these matrices by  selecting $G_1$, $G_2$, is
that the pair $(A_p, B_2[W_{11}~W_{12}])$ is controllable and
the pair $(A_p,\left[\begin{array}{c} S_{11} \\
    S_{21}\end{array}\right]B_1^\dagger)$ 
is observable; the latter condition is equivalent to the controllability of
the pair   
$(A_p^\dagger,B_1\left[\begin{array}{cc} S_{11}^\dagger & S_{21}^\dagger\end{array}\right])$. 
Indeed, these controllability and observability conditions
imply that  
$(A_p-B_2(W_{11}S_{11}+W_{12}S_{21})B_1^\dagger, \left[\begin{array}{c}
    S_{11} \\ S_{21}\end{array}\right]B_1^\dagger)$ is
observable and 
$(A_p-B_2(W_{11}S_{11}+W_{12}S_{21})B_1^\dagger, B_2[W_{11}~W_{12}])$ is
controllable. Therefore, if $(A_p, B_2[W_{11}~W_{12}])$ and
$(A_p^\dagger,B_1\left[\begin{array}{cc} S_{11}^\dagger &
    S_{21}^\dagger\end{array}\right])$ are controllable, 
one can always select $G_1$ and $G_2$ so that the matrices $\hat A$,
$\check A$ have a required eigenvalue distribution. 
Thus the conditions of Theorem~\ref{DFS.synthesis.theorem} boil down to solving
a simultaneous pole assignment problem  under an LMI constraint. 

We next demonstrate that our pole assignment problem captured quantum
plant-controller DFS architectures considered in~\cite{Pan2016,Yamamoto2014}.   

\subsection{Special case 1: DFS synthesis using a coherent observer~\cite{Pan2016}} 
\begin{figure}[t]
\begin{center}
\psfrag{Quantum}{\hspace{-2ex}Quantum}
\psfrag{system}{\hspace{-2ex}system}
\psfrag{controller}{\hspace{-2ex}observer}
\psfrag{w}{$w$}
\psfrag{u}{$u$}
\psfrag{y}{$y$}
\psfrag{f}{$f$}
\psfrag{s}{$s$}
\psfrag{r}{$r$}
\psfrag{z}{$z$}
\psfrag{v}{$v$}
\psfrag{y'}{$y'$}
\psfrag{z'}{$z'$}
\psfrag{u'}{$u'$}
\psfrag{u1'}{$\tilde u'$}
\includegraphics[width=0.4\columnwidth]{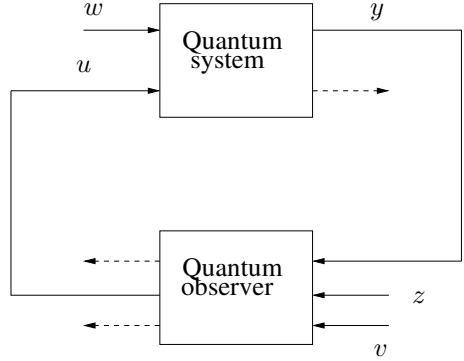}
\caption{Coherent plant-observer network considered in~\cite{Pan2016}.}
\label{fig:ccc}
\end{center}
\end{figure}

In~\cite{Pan2016}, the DFS synthesis was carried out using a quantum
analog of the Luenberger observer for a class of linear annihilation only
systems with a Hamiltonian and a coupling operator described in
(\ref{ham.p}), (\ref{capl.p}); see Fig.~\ref{fig:ccc}. This controller 
structure is a special case of the architecture in
Fig.~\ref{fig:obs-dfs1}, when the two channels $w$ and $f$ are combined as
per (\ref{eq:annihilationplant.dyna}), and 
\[
S=\left[\begin{array}{cc} I & 0 \\ 0 & I
  \end{array}\right], \quad W=\left[\begin{array}{cc} 0 & I \\ I & 0
  \end{array}\right]
\]

With this choice of $S$ and $W$, we have from (\ref{Ao})
\begin{eqnarray}
  \label{Ao.C1}
  A_c=A_p +G_1B_1^\dagger -B_2G_2^\dagger 
\end{eqnarray}

\begin{Col}\label{orig.case}
Suppose the pair $(A_p,C_p)$ is observable and the pair $(A_p,B_2)$ is
controllable. Let $G_1$, $G_2$ be such that
\begin{enumerate}[(a)]
\item
The following linear matrix inequality (LMI) is satisfied
\begin{eqnarray}
  \label{LMI.C1}
\left[\begin{array}{ccc}
R & G_1 & G_2\\
G_1^\dagger & -I & 0 \\
G_2^\dagger & 0 & -I
\end{array}
\right]\le 0,
\end{eqnarray}
where
\begin{eqnarray}
\lefteqn{R=
-B_1B_1^\dagger -B_2B_2^\dagger} && \nonumber \\
&& 
+G_1B_1^\dagger + B_1G_1^\dagger
-B_2G_2^\dagger
-G_2B_2^\dagger
\end{eqnarray}

\item
The matrices 
\begin{eqnarray}
\hat A&=&A_p-B_2G_2^\dagger,   \label{eq:Ahat.C1} \\
\check A&=&A_p+G_1B_1^\dagger \label{eq:Acheck.C1}
\end{eqnarray}
have all eigenvalues on the imaginary axis or in the left half-plane of the
complex plane, with at least one of them having eigenvalues on the
imaginary axis.  
\end{enumerate}
Then the closed loop system admits a DFS. 
\end{Col}

\emph{Proof: }
Via the Schur complement, condition (\ref{LMI.C1}) 
is equivalent to the condition
\begin{eqnarray*}
\lefteqn{-B_1B_1^\dagger -B_2B_2^\dagger +G_1B_1^\dagger + B_1G_1^\dagger} && \\
&&-B_2G_2^\dagger -G_2B_2^\dagger +G_1G_1^\dagger + G_2G_2^\dagger \le 0.
\end{eqnarray*}
This ensures that 
\[
A_c+A_c^\dagger + G_1G_1^\dagger + G_2G_2^\dagger \le 0 .
\]
Therefore, one can find $G_3$ such that the controller is physically
realizable. The claim then follows from
Theorem~\ref{DFS.synthesis.theorem}.

\hfill$\Box$

\subsection{Special Case 2: Coherent feedback DFS generation model
  from~\cite{Yamamoto2014}} 

\begin{figure}[t]
\begin{center}
\psfrag{Quantum}{\hspace{-2ex}Quantum}
\psfrag{system}{\hspace{-2ex}system}
\psfrag{controller}{\hspace{-2ex}controller}
\psfrag{w}{$w$}
\psfrag{u}{$u$}
\psfrag{y}{$y$}
\includegraphics[width=0.4\columnwidth]{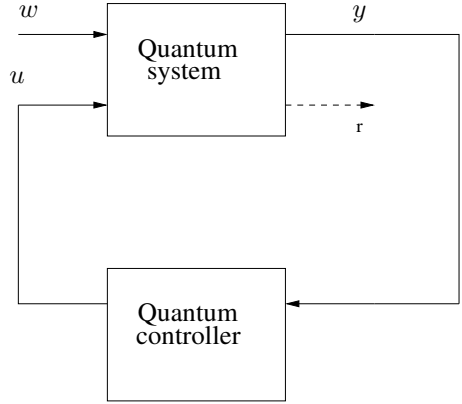}
\caption{Special Case 2: Coherent feedback network for DFS generation
  considered in~\cite{Yamamoto2014}.}
\label{fig:obs-dfs2}
\end{center}
\end{figure}

Consider a system of Fig.~\ref{fig:obs-dfs1} in which 
$S=I$, $W=I$, and let $G_2=0$, $G_3=0$. 
This corresponds to the system shown in Fig.~\ref{fig:obs-dfs2};
which was considered in~\cite{Yamamoto2014}.   In this case, the controller matrix
becomes 
\begin{eqnarray}
  \label{Ao.C2}
  A_c=A_p-B_2B_1^\dagger +G_1B_1^\dagger -B_2G_1^\dagger.
\end{eqnarray}

\begin{Col}\label{Naoki}
Suppose the pair $(A_p,C_p)$ is observable and the pair $(A_p,B_2)$ is
controllable. Let $G_1$ be such that
\begin{enumerate}[(a)]
\item
The following equation is satisfied
\begin{eqnarray}
  \label{LMI.C2}
-(B_1+B_2)(B_1+B_2)^\dagger+G_1(B_1-B_2)^\dagger \nonumber \\
+(B_1-B_2)G_1^\dagger+G_1G_1^\dagger =0;
\end{eqnarray}
\item
The matrices 
\begin{eqnarray}
\hat A&=&A_p-B_2B_1^\dagger -B_2G_1^\dagger, \label{eq:Ahat.C2} \\
\check A&=&A_p-B_2B_1^\dagger+G_1B_1^\dagger \label{eq:Acheck.C2}
\end{eqnarray}
have all eigenvalues on the imaginary axis 
or in the left half-plane of the
complex plane, with at least one of them having eigenvalues on the
imaginary axis.   
\end{enumerate}
Then the closed loop system admits a DFS. 
\end{Col}

\emph{Proof: }
Condition (\ref{LMI.C2}) ensures that 
\[
A_c+A_c^\dagger + G_1G_1^\dagger = 0 .
\]

Next, $\hat A$ and  $\check A$ have eigenvalues on the imaginary axis or in
the open left half-plane, hence the statement of the corollary follows from
Theorem~\ref{DFS.synthesis.theorem}.
\hfill$\Box$

\section{Examples}
\label{sec:ex}

\subsection{Example~1}
To illustrate the DFS synthesis procedure developed in the
previous section, consider a system consisting of two optical
cavities interconnected as shown in Fig.~\ref{fig:ccc}. The system is
similar to those considered in~\cite{Nurdin2015}.

The cavity to be controlled is described by
equation~(\ref{eq:annihilationplant.dyna}), with all matrices becoming
complex numbers  
\begin{eqnarray}
A_p &=& -iM-\frac{\kappa_1+\kappa_2}{2},\quad B_1 = - \sqrt{\kappa_1},\quad B_2 =  - \sqrt{\kappa_2},  \nonumber \\
C_p &=& -B_1^*= \sqrt{\kappa_1}.
\end{eqnarray}
Here, $\kappa_1$, $\kappa_2$ are real nonnegative numbers, characterizing the
strength of the couplings between the cavity and the input fields $w$ and $u$,
respectively, and $M$ characterizes the Hamiltonian of the cavity. 

Clearly, the pair $(A_p,C_p)$ is observable and the pair $(A_p,B_2)$ is
controllable, therefore the optical cavity cannot have a DFS unless the
cavity is lossless. To synthesize a DFS, let us connect this cavity to another
optical cavity with the same Hamiltonian, as shown in Fig.~\ref{fig:ccc}.
This corresponds to letting the controller have the coefficients   
 \begin{eqnarray}
A_c &=& -iM-\frac{\kappa_3+\kappa_4}{2},\quad G_1 =  -\sqrt{\kappa_3} ,\quad G_2 =  - \sqrt{\kappa_4},  \nonumber \\
K &=&  \sqrt{\kappa_4} , \quad G_3=0.
\end{eqnarray}
and letting the scattering matrices $S$ and $W$ be 
\begin{equation}
S = \left[\begin{matrix} 1 & 0  \\ 0 & 1  \end{matrix} \right], \quad W = \left[\begin{matrix} 0 & 1  \\ 1 & 0  \end{matrix} \right].
\end{equation}
We now apply Corollary~\ref{orig.case} to show that the parameters
$\kappa_3, \kappa_4$ for the controller cavity can be chosen so that the
two-cavity system has a DFS. It is readily verified that the matrices
$\hat{A}$ and $\check{A}$ in (\ref{eq:Ahat.C1}),~(\ref{eq:Acheck.C1})
reduce to   
\begin{eqnarray}
\hat{A} &=& -iM -\frac{\kappa_1+\kappa_2}{2} -\sqrt{\kappa_2\kappa_4},\\
\check{A} &=& -iM -\frac{\kappa_1+\kappa_2}{2} +\sqrt{\kappa_1\kappa_3}.
\end{eqnarray}

From Corollary~\ref{orig.case}, we need either $\hat{A}$ or $\check{A}$ to
have  poles on the imaginary axis in order to create a DFS within the
closed-loop system. Clearly, for the two-cavity system under consideration
this can only be achieved by placing the pole of $\check{A}$ at the
origin. For this, the coupling rate $\kappa_3$ of the controller must
be set to 
\begin{equation}
\kappa_3 = \frac{(\kappa_1+\kappa_2)^2}{4\kappa_1}.
\label{eq:k3}
\end{equation}
Also we must satisfy the LMI condition (\ref{LMI.C1}). The matrix $R$ in
this example reduces to
\begin{eqnarray}
R = -\kappa_1 -\kappa_2+2\sqrt{\kappa_1\kappa_3}-2\sqrt{\kappa_2\kappa_4}.
\label{eq:R}
\end{eqnarray}
Hence, using  (\ref{eq:k3}) and (\ref{eq:R}), the LMI condition
(\ref{LMI.C1}) reduces to the two following inequalities:
\begin{equation}
\text{Re}\left[-\kappa_1 - 2 \kappa_1 \sqrt{\kappa_2 \kappa_4} \pm \sqrt{D}\right]\leq 0,
\label{eq:k4}
\end{equation}
where
\begin{eqnarray*}
D=\kappa_1^2 + \kappa_1^3 + 2 \kappa_1^2 \kappa_2 + \kappa_1 \kappa_2^2 +
4 \kappa_1^2 \kappa_4 \\
+ 4 \kappa_1^2 \kappa_2 \kappa_4 - 4 \kappa_1^2 \sqrt{\kappa_2 \kappa_4}.
\end{eqnarray*}
The inequality (\ref{eq:k4}) is the only constraint for the remaining
coupling parameter $\kappa_4$ to be determined. Notice that there is an
obvious solution to this inequality in the case where $\kappa_1 = \kappa_2
= \kappa$. The solution is $\kappa_3 = \kappa_4 = \kappa$ which satisfies
both (\ref{eq:k3}) and  (\ref{eq:k4}).  

The above calculations demonstrate that by placing the pole of the
controller on the imaginary axis, one can effectively create a DF mode
which did not exist in the original system. This fact has been
established previously in~\cite{Nurdin2015} by calculating the system poles,
whereas we have arrived at this conclusion from a more general
Corollary~\ref{orig.case}, as a special case.

\subsection{Example 2}
\begin{figure}[t]
\begin{center}
\psfrag{Quantum}{\hspace{-2ex}Quantum}
\psfrag{system}{\hspace{-2ex}system}
\psfrag{controller}{\hspace{-2ex}controller}
\psfrag{w}{$w$}
\psfrag{u}{$u$}
\psfrag{y}{$y$}
\psfrag{y}{$y$}
\psfrag{r}{$r$}
\psfrag{k1}{$\kappa_1$}
\psfrag{k2}{$\kappa_2$}
\psfrag{k3}{$\kappa_3$}
\psfrag{k4}{$\kappa_4$}
\psfrag{Cavity 1}{Cavity 1}
\psfrag{Cavity 2}{Cavity 2}
\includegraphics[width=0.4\columnwidth]{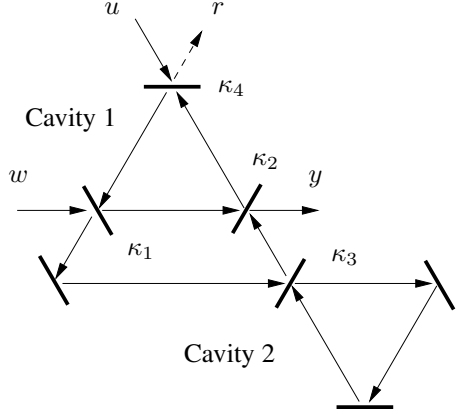}
\caption{The two-cavity system for Example~2.}
\label{2cavity}
\end{center}
\end{figure}

We now present an example in which, the DFS is
created which is shared between the controlled system and the
controller. The controlled system in this example consists of two cavities
as shown in Fig.~\ref{2cavity}.

Denote the matrices associated of the Hamiltonians corresponding to the
each cavity internal dynamics as $M_1$, $M_2$. Also for the
convenience of notation, define the complex numbers 
\[
\gamma_j=\sqrt{\kappa_j}, \quad j=1,\ldots, 4, 
\]  
associated with the coupling strengths within the cavities. All four constants
are assumed to be nonzero.

Then the equations governing the dynamics of the two-cavity system have the
form of (\ref{eq:annihilationplant.dyna}) with 
\begin{eqnarray}
  \label{eq:ex2.1}
  A_p&=&\left[
    \begin{array}{cc}
      -\left(iM_1+\frac{|\gamma_1|^2+|\gamma_2|^2}{2}+\gamma_1^*\gamma_2\right) &
      -\gamma_2\gamma_3^* \\
-\gamma_1^*\gamma_3 & -\left(iM_2+\frac{|\gamma_3|^2}{2}\right)
    \end{array}
\right], \nonumber \\ B_1&=&\left[
    \begin{array}{c}
-(\gamma_1+\gamma_2) \\ -\gamma_3
\end{array}\right], \quad B_2=\left[
    \begin{array}{c} -\gamma_4 \\ 0\end{array}\right], \nonumber \\
C_p&=&-B_1^\dagger = \left[\begin{array}{cc}\gamma_1+\gamma_2 & \gamma_3\end{array}\right].
\end{eqnarray}

To verify observability of the pair $(A_p,C_p)$, we observe that
\begin{eqnarray*}
\det\left[\begin{array}{c}C_p \\ C_pA_p
  \end{array}
\right]&=&\frac{1}{2}\gamma_3^*\left(|\gamma_3|^2(\gamma_1^*-\gamma_2^*)
\right.\\
&&\left. +(\gamma_1^*+\gamma_2^*)(|\gamma_1|^2-|\gamma_2|^2)\right.\\
&&\left. +2i(\gamma_1^*+\gamma_2^*)(M_1-M_2)\right).
\end{eqnarray*}
Suppose $\gamma_1=-\gamma_2$, then
$\det\left[\begin{array}{c}C_p\\C_pA_p
  \end{array}\right]=\gamma_3^*\gamma_1^*|\gamma_3|^2$, and we conclude
that the matrix $\left[\begin{array}{c}C_p\\C_pA_p
  \end{array}\right]$ is full rank. This implies that in the case $\gamma_1=-\gamma_2$, the pair $(A_p,C_p)$
is observable. Also, the pair $(A_p,B_2)$ is controllable, since  
\[
\det\left[B_2~A_pB_2\right]=-\gamma_4^2\gamma_3\gamma_1^* \neq 0.
\]  
These observations allow us to apply Corollary~\ref{orig.case} to construct
a DFS by interconnecting the two-cavity system with a coherent quantum
observer, which we now construct.

For simplicity, choose 
\[
G_1=\left[\begin{array}{c}g_1 \\ 0 \end{array}\right], \quad
G_2=\left[\begin{array}{c}0 \\ g_2 \end{array}\right].
\]
With this choice of $G_1$ and $G_2$ and under the condition
$\gamma_1=-\gamma_2$, the matrices $\hat
A=A_p-B_2G_2^\dagger$ and $\check A=A_p+G_1B_1^\dagger$ take the form
\begin{eqnarray*}
  \hat A&=&\left[\begin{array}{cc}
      -iM_1 &
      -\gamma_2\gamma_3^*+\gamma_4g_2^* \\
-\gamma_1^*\gamma_3 & -\left(iM_2+\frac{|\gamma_3|^2}{2}\right)
  \end{array}\right], \\
  \check A&=&\left[\begin{array}{cc}
      -iM_1
    &
      -\gamma_2\gamma_3^* -g_1\gamma_3^* \\ 
-\gamma_1^*\gamma_3 & -\left(iM_2+\frac{|\gamma_3|^2}{2}\right)
  \end{array}\right].
\end{eqnarray*}
Letting $g_2=\frac{\gamma_2^*\gamma_3}{\gamma_4^*}$, $g_1=-\gamma_2$ allows
us to conclude that each of the matrices $\hat A$ and $\check A$ have one
imaginary eigenvalue and one eigenvalue with negative real part,
$-\left(iM_2+\frac{|\gamma_3|^2}{2}\right)$. 

It remains to show that the LMI condition (\ref{LMI.C1}) is satisfied in
this example. Noting that with the above choice of $g_1$, $g_2$,
\[
R= \left[\begin{array}{cc} -|\gamma_4|^2 & 2\gamma_2\gamma_3^* \\
 2\gamma_2^*\gamma_3 & -|\gamma_3|^2
\end{array}\right],
\]
$R<0$ holds provided $|\gamma_4|^2>4|\gamma_2|^2$. Next, the LMI
(\ref{LMI.C1}) in this example requires that
\begin{eqnarray}
\left[\begin{array}{cccc}
-|\gamma_4|^2 & 2\gamma_2\gamma_3^* & -\gamma_2 & 0   \\ 
 2\gamma_2^*\gamma_3 & -|\gamma_3|^2 & 0  &
 \frac{\gamma_2^*\gamma_3}{\gamma_4^*} \\
-\gamma_2^* & 0 & -1 & 0 \\
0 & \frac{\gamma_2\gamma_3^*}{\gamma_4}  & 0 & -1 
\end{array}\right]<0
\end{eqnarray}
Using the Schur complement, this requirement is equivalent to 
\[
\left[\begin{array}{cc}
-|\gamma_4|^2 & 2\gamma_2\gamma_3^* \\
 2\gamma_2^*\gamma_3 & -|\gamma_3|^2\end{array}\right]
+\left[\begin{array}{cc}
|\gamma_2|^2 & 0 \\ 0 & \frac{|\gamma_2|^2|\gamma_3|^2}{|\gamma_4|^2} 
\end{array}\right]<0.
\] 
The latter condition holds when $|\gamma_4|^2>(3+\sqrt{10})|\gamma_2|^2$.

\section{Concluding remarks}
\label{sec:con}
In this paper, we have proposed a general coherent
quantum  controller synthesis procedure for generating decoherence free
subspaces in quantum systems. Decoherence free components capable of
storing quantum information are regarded to be essential for quantum computation
and communication, as quantum memory elements~\cite{Nurdin2015}. When the
feedback loop is in the DFS configuration, the DFS mode is
`protected', which also means that to access dynamics of that mode, the
system must be augmented with a mechanism to dynamically change the
feedback configuration in order to bring the system in and out of the `DF
state'. E.g., from the above examples, we see that adjusting the
values of coupling strengths is one possibility to achieve
this. However, this approach is only applicable for experimental systems
which have tunable coupling devices available, such as an optical waveguide
or a microwave superconducting cavity. Another viable approach for the
systems in those examples would be changing the loop configuration by using
optical switches to either break the
feedback loop or form an additional feedback
connection, i.e., form a double-pass feedback loop; the latter is
essentially the approach presented in~\cite{Nurdin2015}. Our future work will
consider these approaches in greater detail, to obtain general dynamical
reading and writing procedures augmenting our general results in a fashion
similar to how this has been done in~\cite{Nurdin2015} for optical cavity
systems.

\end{document}